\documentclass[12pt]{iopart}

\usepackage{iopams}  
\usepackage{cite}
\usepackage{algorithmic}
\usepackage{graphicx}
\usepackage{textcomp}
\def\BibTeX{{\rm B\kern-.05em{\sc i\kern-.025em b}\kern-.08em
    T\kern-.1667em\lower.7ex\hbox{E}\kern-.125emX}}
\usepackage{orcidlink}
\usepackage{siunitx}

\usepackage{support-caption} 
\usepackage{subcaption}

\usepackage{soul}

\DeclareUnicodeCharacter{2212}{-}
\usepackage{hyperref} 

\begin{document}

\title[OPM with High Spatial Resolution Guide]{Optically Pumped Magnetometer with High Spatial Resolution Magnetic Guide for the Detection of Magnetic Droplets in a Microfluidic Channel}

\author{Marc Jofre$^{1,2}$\orcidlink{0000-0002-8912-6595}, Jordi Romeu$^{1}$\orcidlink{0000-0003-0197-5961} and Luis Jofre-Roca$^1$\orcidlink{0000-0002-0547-901X}}

\address{1. Dept. Signal Theory and Communications, Technical University of Catalonia, Barcelona 08034, Spain}
\address{2. Dept. Research and Innovation, Fundaci\'o Privada Hospital Asil de Granollers, Granollers 08402, Spain}
\ead{marc.jofre@upc.edu}
\vspace{10pt}
\begin{indented}
\item[]January 2023
\end{indented}

\begin{abstract}
Quantum sensors provide unprecedented magnetic field detection sensitivities, enabling these to extend the common magnetometry range of applications and environments of operation. In this framework, many applications also require high spatial resolution magnetic measurements for biomedical research, environmental monitoring and industrial production. In this regard, Optically Pumped Magnetometers (OPMs) are considered as prominent candidates, but are impaired in size with micrometer scale magnetic particles, e.g. magnetic droplets. In order to address this limitation, here we study the effects of adding a micrometer-to-millimeter magnetic guide to a miniature OPM. This device is applied to detect Fe3O4 magnetic droplets flowing at rates up to $25$ drop./s in a microfluidic channel. The computed spatial resolution is \SI{300}{\micro\meter} and the measured SNR is larger than \SI{15}{\decibel} for the different sizes of considered magnetic droplets.
\end{abstract}

%
\vspace{2pc}

\section{Introduction}\label{sec:introduction}
Recent advances in science and technology are producing a wide range of bioparticle interaction and sensing possibilities, eventually even leading to communication with organisms, that will drastically change the way we link to them. The fusion of fields like nano, bio, and microfluidics technologies is one pertinent example. These technological combinations have culminated in various brands and concepts that are currently used in a variety of research fields, such as Lab-on-a-chip (LoC)~\cite{abgrall_chip_2007}. For instance, improved miniaturization techniques enable the integration of a full system on a chip (SoC)~\cite{shah_fully_2018,tierney_optically_2019,sulai_characterizing_2019} with sensing, processing, and wireless interactive functions~\cite{gi_optoelectrofluidic_2017}, including microfluidic and detection subsystems~\cite{coleman_quantum_2021,budker_optical_2007,etcheverry_high_2017,zarrinkhat_experimental_2022}.

When trying to approach the contact-less environment for basic particle interaction, and in particular for droplets containing magnetic particles, mimicking magnetic signals from bioparticles (e.g., action potentials from cells)~\cite{shang_droplet-templated_2021,martino_droplet-based_2016,kim_microfluidic_2009}, the combination of magnetic sensors with microfluidics has proven to be a very interesting pairing for detecting and controlling the functionality of droplets through the interplay with fluid streams~\cite{bougas_nondestructive_2018,nguyen_fundamentals_2006,carlo_inertial_2009,convery_30_2019}; performing these functions without mechanical moving parts, and with little noise generation in the process~\cite{jofre_wireless_2021,russom_differential_2009,palacios_superheterodyne_2022}.

In this direction, quantum technologies, jointly with nanotechnologies, have led to a new field of physics and engineering aimed at translating the effects of quantum physics into practical applications. Using solid-state quantum systems, atoms, ions, molecules, and photonic circuitry, one can design macroscopic but intrinsically quantum devices used for receiving dim signals~\cite{degen_quantum_2017,gerginov_prospects_2017,fu_sensitive_2020,amoros-binefa_noisy_2021,mitchell_scale-invariant_2020,aharon_fully_2016,fancher_rydberg_2021,acin_quantum_2018,wong_scqis_2022}, i.e. Optically Pumped Magnetometers (OPMs).

Moreover, materials with high magnetic permeability tend to collect magnetic flux lines. This property can lead to a local magnetic field amplification and enhanced spatial resolution in the gap between two high permeability structures or flux concentrators. These have previously been combined with a wide variety of magnetic sensor technologies~\cite{kim_ultra-sensitive_2016,griffith_miniature_2009,put_zero_2021,soares_go_2019}.

Nevertheless, limitations with the detection of magnetic droplets in microfluidic channels are still present. On the one hand, there is a dimension mismatch when interfacing between magnetic droplets and atomic magnetometers, because the latter is relatively large compared to micro-sized droplets. This unpairing of dimensions causes the lack of spatial resolution to individually interrogate droplets. On the other hand, magnetic detection systems suffer from loss of signal sensitivity and integrity due to intricate coupling structures and stray fields, the latter from the measuring system as well as from the environment. 

Therefore, in this work it is proposed a magnetic guide on an OPM to allow high spatial resolution, while maintaining the sensitivity, for detecting magnetic droplets in a microfluidic channel. The magnetic guide is based on a magnetic hose structure that allows efficiently transferring the magnetic signal at a distance~\cite{navau_long-distance_2014}, which at the same time is convenient to match the size gap between magnetic droplets in microchannels and OPMs.

Correspondingly, the objectives of this work are to: 1) derive an OPM with high spatial resolution magnetic guide framework for detecting magnetic droplets in a microfluidic channel; 2) provide calculations and experimental demonstration of the capabilities achieved in combination with a test system; and 3) discuss the results at detection rates up to $25$ droplets/s (drop./s). In this regard, this article is organized as follows. First, Section \ref{SecDeviceOpeartion} describes the constructed system and its principle of operation. Then, an analysis of the performance in terms of spatial resolution and sensitivity is provided in Sections \ref{SecSpatialResolutionDiscussion} and \ref{SecTheoreticalDescription}, respectively. Experimental demonstrations of the system and discussion of the results are provided in Section \ref{SecExperimentalResults}. Finally, Section \ref{SecConclusion} summarizes the work and provides conclusions.

\section{Device construction and principle of operation}\label{SecDeviceOpeartion}
\begin{figure}[t!]
     \centering
     \begin{subfigure}[b]{0.7\columnwidth}
         \centering
         \includegraphics[width=\columnwidth]{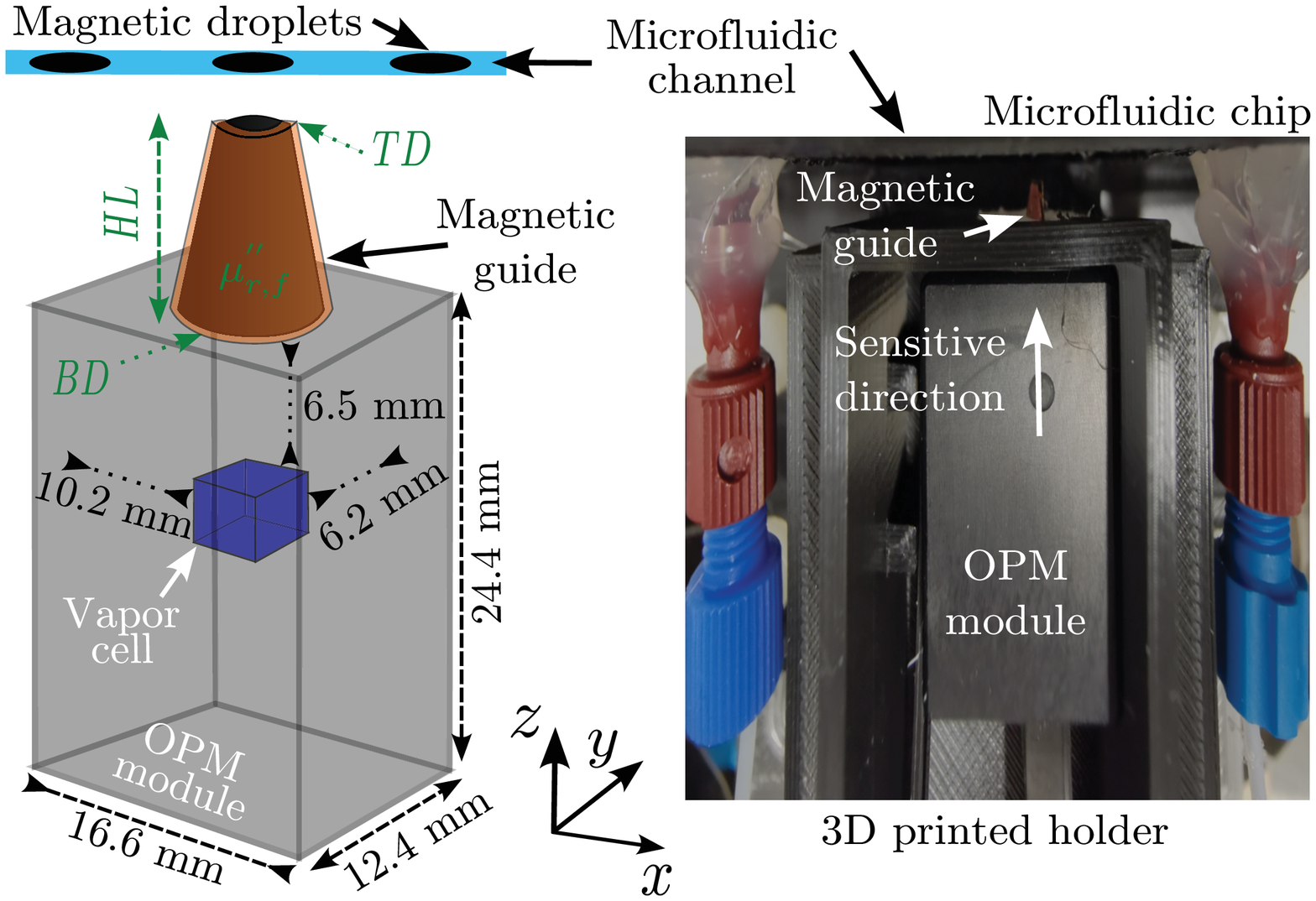}
         \caption{}
         \label{fig:ConstructionDevice1}
     \end{subfigure}
     \hfill
     \begin{subfigure}[b]{0.75\columnwidth}
         \centering
         \includegraphics[width=\columnwidth]{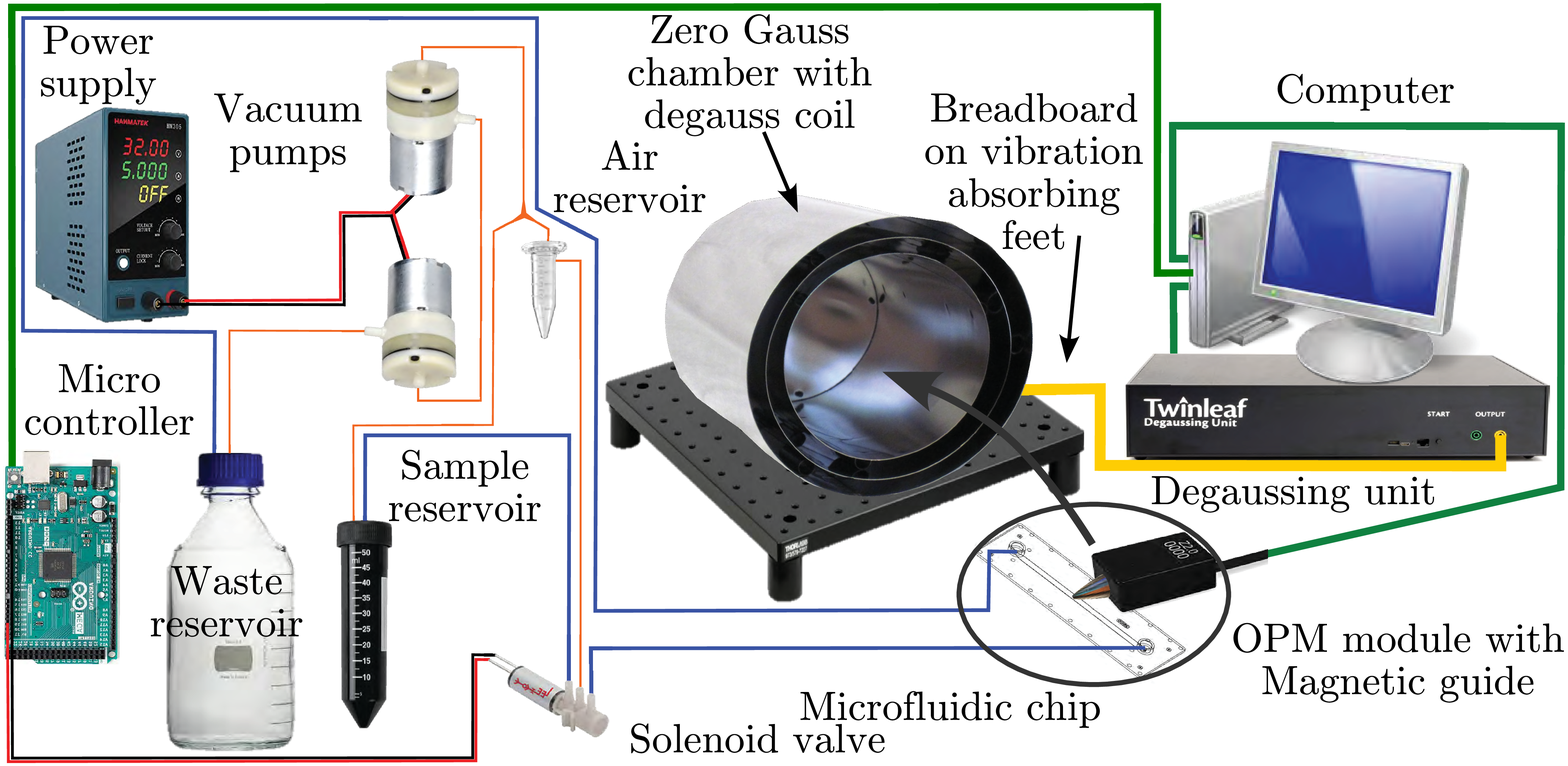}
         \caption{}
         \label{fig:MagneticMicrofluidicsSetup1}
     \end{subfigure}
    \caption{Schematics of (a) the OPM with conical magnetic guide (OPM-MG) aligned to the microchannel and (b) the test system.}
    \label{fig:SchematicDesingSetup}
\end{figure}

The device uses an OPM with an attached cone-shaped magnetic guide (OPM-MG) for detecting magnetic droplets in a microfluidic channel, as shown in Fig. \ref{fig:ConstructionDevice1}, providing high spatial resolution while preserving magnetic sensitivity.

The OPM-MG is assembled inside a Zero Gauss chamber and it is further composed of a microfluidic handling platform for generating magnetic droplets, and operation/reading electronics controlled by a computer. Figure \ref{fig:MagneticMicrofluidicsSetup1} details the different elements of the test system and its connections (both electrical and fluidics) with the following color code: liquid tubes (blue), air lines (orange), data cables (green), degauss coil current line (yellow) and voltage lines (red/black). 

\subsection{Magnetic guide}\label{DescriptionMG}
The magnetic guide (MG) is based on transformation optics operating in the subwavelength regime which efficiently routes magnetic fields~\cite{navau_long-distance_2014}. To detect magnetic droplets in a microfluidic channel, it is convenient to match the microchannel size to the size of the OPM with a cone-shaped MG~\cite{kondo_evaluation_2003}. As detailed in Fig. \ref{fig:ConstructionDevice1}, the MG is made of a ferrite core of \SI{5}{\milli\meter} in length ($HL$), \SI{3}{\milli\meter} base diameter ($BD$), and \SI{0.2}{\milli\meter} top diameter. The core is covered by a superconductor-strip-tape (Theva Pro-Line tape 4121 HTS). The tape is \SI{12}{\milli\meter} wide and \SI{80}{\micro\meter} thick, and it is cut accordingly to cover the lateral sides of the core of the MG. The superconducting tape was tightly glued to the core in order to prevent flux leakage. 

\subsection{OPM module}\label{OPMmodule}
Optically Pump Magnetometers (OPMs) rely on the manipulation of a quantum property know as spin (a property that underlies a particle's magnetic moment and therefore its response to a magnetic field), first described in the 1920s and 1930s~\cite{rabi_new_1938,gerlach_experimentelle_1922}. In particular, OPMs are fundamentally based on the manipulation of the atomic (i.e. both nuclear and electron) spin~\cite{bloch_nuclear_1946,bell_optical_1957,groeger_comparison_2005,kitching_microfabricated_2008,zhang_improving_2018}.

The OPM device used is a $\mathrm{cm}$-size zero-field QuSpin OPM with $3$ × $3$ × \SI{3}{\milli\meter\cubed} Rb vapor cell operating in Spin-Exchange Relaxation Free (SERF) regime \cite{coleman_quantum_2021,dupont-roc_detection_1969,shah_compact_2013,allred_high-sensitivity_2022}, as shown in Fig. \ref{fig:SchematicDesingSetup}. The typical sensitivity is \SI{10}{\femto\tesla/\sqrt\Hz} in single-axis mode and \SI{3}-\SI{100}{\hertz} operating frequency range.

In particular, the OPM is an absolute zero field magnetometer and operates best ($<1\%$ non-linearity) when the background field is \SI{<1}{\nano\tesla}. The OPM has three compensation coil pairs that surround the vapor cell like a cube (\SI{\sim5}x\SI{5}{\milli\meter} surface with $5$ turns in each coil). The coils do not create a completely homogeneous magnetic field, so they can not be operated to produce very large fields (they ought to operate at \SI{<200}{\nano\tesla} to not create strong gradients causing sensitivity losses).

In addition, when calibrating, a known magnetic field is momentarily applied with the internal coils to calibrate the magnetometer response. The calibration process yields a calibration number which is multiplied with the raw output in a way that scales the analog output to be \SI{2.7}{\volt/\nano\tesla} (in default mode, $1$x) and the digital output to be represented in pT. In particular, the calibration number algorithm is proprietary of the manufacturer and it is capped at $15.9$. If the magnetometer is left uncalibrated, the calibration factor is set at one. For this work, the quality of calibration achieved was $0.76$ in the $z$-axis, hence the magnetometer is relatively healthy ($<1$).

\subsection{Residual magnetic field}
To remove the ambient magnetic DC field and noise, to $<10$’s of nT, the OPM-MG is located inside a three-layer closed mu-metal cylindrical Zero Gauss chamber (ZG-206 W/Degauss, Magnetic Shield Corp.) with inside diameter (inside depth) of \SI{152}{\milli\meter} (\SI{381}{\milli\meter}) and with a degauss coil installed and operated with a degaussing (DG) unit (digital unit, Twinleaf).

The DG unit is used to automate the process of degaussing the magnetic shield and the magnetic components inside the chamber, by applying an oscillating magnetic field that drives the material into saturation and then slowly ramping the oscillation amplitude down to zero. The DG unit is operated with $5$ sequences of \SI{100}{\milli\ampere} \SI{50}{\hertz} degaussing pulses of \SI{5}{\second} to demagnetize the ferrite elements. Eventually, if the demagnetization is not sufficient, the internal lock-in process used by the OPM, both in order to detect the magnetic field levels and to auto compensate less than nT fields, generates a message in order for the operator to further reduce residual magnetic fields or magnetic field gradient inhomogeneities, as discussed in detail in Fig. \ref{fig:TheorAnalysisSensitivityFreqDifferentInho} in subsection \ref{SecGradientSensLoss}.

The maximum background magnetic field amplitude that the OPM can operate in is $\sim$\SI{50}{\nano\tesla} due to the internal coils. The first limitation of the residual field is set by the signal-to-noise ratio (SNR) of the digital-to-analog converters (DACs) that control the compensation field. The second limitation is the field inhomogeneity created by the internal coils, which broadens the zero-field resonance and consequently lowers the sensor performance. 

\subsection{Mechanical structure}
\begin{figure}[t!]
     \centering
     \begin{subfigure}[b]{0.69\columnwidth}
         \centering
         \includegraphics[width=\columnwidth]{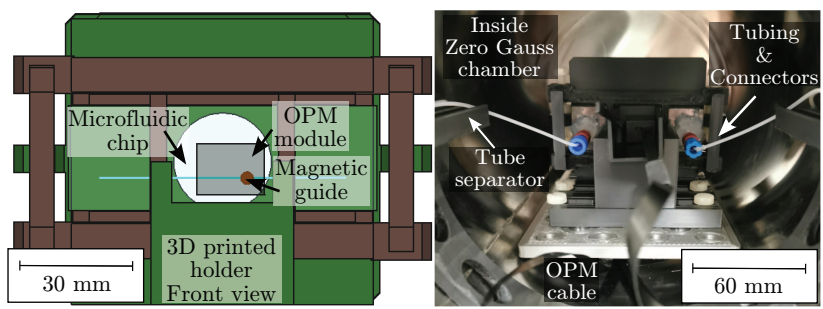}
         \caption{}
         \label{fig:MountSetup1}
     \end{subfigure}
     \begin{subfigure}[b]{0.69\columnwidth}
         \centering
         \includegraphics[width=\columnwidth]{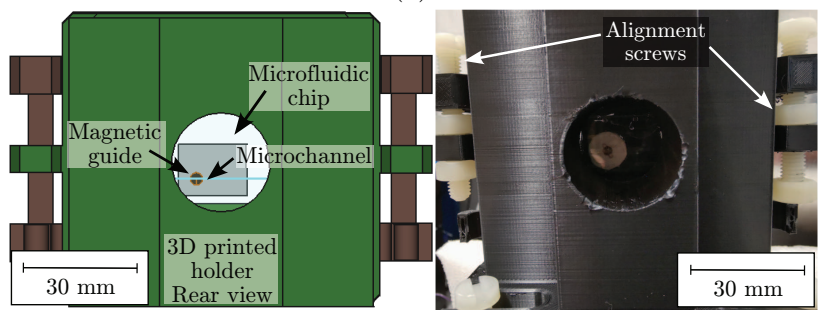}
         \caption{}
         \label{fig:MountSetup2}
     \end{subfigure}
     \begin{subfigure}[b]{0.69\columnwidth}
         \centering
         \includegraphics[width=\columnwidth]{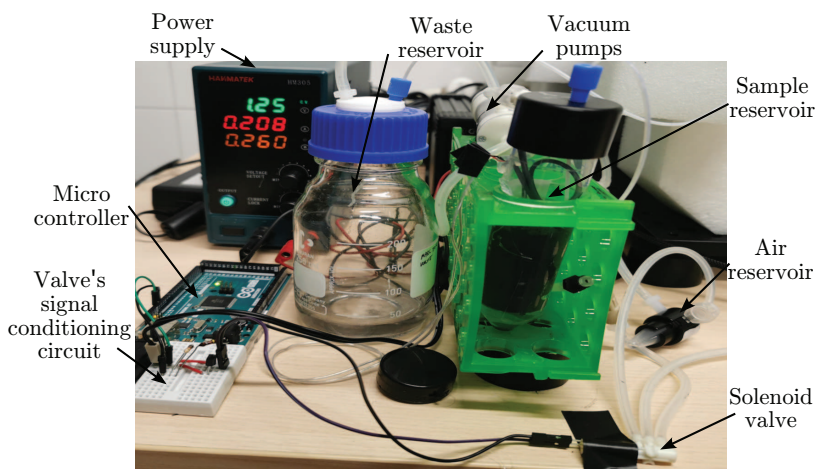}
         \caption{}
         \label{fig:MountSetup3}
     \end{subfigure}
    \caption{Schematics and prototype of the assembled mounting structure to hold the OPM-MG and the microfluidic channel chip. (a) Inside the Zero Gauss chamber, (b) close view of the alignment system to the microchannel, and (c) the microfluidic system assembly.}
    \label{fig:ExpSetup}
\end{figure}

A polymer polylactic acid (PLA) 3D printed mechanical holder, with \SI{0.1}{\milli\meter} resolution, was constructed for the OPM-MG. The holder allows positioning the OPM-MG in the center of the Zero Gauss chamber, placing the microfluidic tubes away from the OPM module (with tube separators). The different cables and tubes are directed towards the in/out opening of the can, as shown in Fig. \ref{fig:MountSetup1}. Conveniently, the holder does not completely cover the OPM module, allowing for a proper temperature convection. Furthermore, the mechanical holder allows the alignment of the tip of the MG to the center of the microfluidic channel, as shown in Fig. \ref{fig:MountSetup2}.

Finally, the Zero Gauss chamber is mounted on a breadboard with absorbing feet to reduce noise due to mechanical vibrations of the surrounding with an isolation of more than \SI{20}{\decibel}.

\subsection{Magnetic droplets generation and microfluidics}\label{SecMagDropMicrofluidics}
The magnetic droplets are composed of tiny (\SI{1}-\SI{20}{\micro\meter}) iron oxides, i.e. magnetite (Fe3O4), at $10^5$ particles/mL solute in water. For the latter magnetite particle sizes and concentration in the microchannel, the \SI{0.01} - {0.2} blockage ratio is low, more than an order of magnitude difference in particle sizes is present and the sample is diluted enough for particles to occupy homogeneously the overall cross section of the microchannel. Whilst the droplets are generated with an ON/OFF actively operated microfluidic valve directly modulated at rates of \SI{5}-\SI{25} drop./s~\cite{galas_active_2009}. Finally, for the latter frequency rates, the range of lengths of the magnetic droplets do not overlap inside the microchannel and the periodic trains of droplets fall within the detection bandwidth of the OPM.

In addition, the microfluidic platform is operated with two voltage-controlled vacuum pumps connected in series (output of pump 1 to the vacuum input of pump 2) and the output is reconnected into the sample holder for continuous resuspension of the magnetite particles, to avoid deposition (since the particle density is higher than that of water). At a system level, the microfluidic elements are interconnected using ID \SI{125}{\micro\meter} PEEK tubing and electrically wired as shown in Fig. \ref{fig:MountSetup3}.

In this way, the magnetic droplets circulate at a flow rate of \SI{4.5}{\micro\liter/\minute} through a \SI{100} x\SI{100}{\micro\meter} square cross-section \SI{56}{\milli\meter}-long microfluidic channel. In consequence, the flow is laminar with a low Reynolds number ($R_e=0.46$). Also, the microfluidic channel selected for the measurements corresponds to a commercially produced microchip with optical-grade-quality polymethyl methacrylate (PMMA). Accordingly, with this microchip the distance from the tip of the MG to the center of the microchannel is \SI{225}{\micro\meter}, corresponding to the tip being in contact with the top face of the microfluidic chip, a \SI{175}{\micro\meter} cover lid thickness and a \SI{50}{\micro\meter} half-height of the microchannel.

\section{Spatial resolution and magnetic flux transfer analysis}\label{SecSpatialResolutionDiscussion}
The MG concentrates and transfers the magnetic field into the OPM. Therefore, the effective magnetic flux density sensed by the OPM is significantly larger than with respect not having the MG.

Considering a magnetic droplet as a point source, which is equivalent to a magnetic dipole, its magnetic field decays very rapidly with distance as $d^3$. Instead, placing the point source at the tip (or top) of a cylindrical MG and the OPM module at its base (or bottom), the magnetic field in the $z$-axis instead follows $B_z = \frac{\mu_0\mu_{r}^{'}m}{2\pi r^2 \left(l_s+L_a\right)}$, where $r$ is the radius of the FM core, $\mu_0$ the vacuum permeability, $\mu_{r}{'}$ the real relative permeability, $l_s$ ($l_a$) the length of the source (air) gap, and $m$ the magnetic moment~\cite{navau_long-distance_2014}. As well, on the $xy$-plane the magnetic field commonly decreases from the center~\cite{tipler_physics_2003}. Therefore, the spatial resolution and magnetic field transfer capabilities can be adjusted by designing the shape and dimension of the MG.

Therefore, in this section we describe the analysis of the interplay of four different parameters relevant to design the cone-shaped MG. The analysis is supported with numerical computations of magnetostatic fields~\cite{gunawan_magnetic-tip_2020}, using cylindrical coordinates, by solving the vector magnetic potential~\cite{hart_analytical_2020}. Furthermore, imposing the boundary condition that the energy entering at the top surface of the MG is the same as the energy exiting at the bottom part~\cite{navau_long-distance_2014}, in the case of zero magnetic losses. In the calculations, it is considered a real relative permeability of the ferrite core of $\mu^{'}_{r,f}=5\cdot 10^4$, and the superconductor tape being perfectly diamagnetic ($\mu^{'}_{SC}=0$). The calculation is performed on a 32 GB RAM and $6$ $i5$-core computer running Jupyter Lab with Python 3.9.10. Finally, the calculation does not take into account the effects of the magnetic field gradient on the OPM because they are negligible for this work, as discussed in Section \ref{SecGradientSensLoss}.

With the above considerations, the calculation allows optimizing the MG in terms of gain and spatial resolution. First, the gain is related to the capacity of the MG to concentrate and transfer the magnetic field along the MG. In this sense, the gain is computed as the signal at the OPM with MG compared to the signal when placing the OPM vapor cell right next to the magnetic point source without MG. Secondly, the spatial resolution is extracted from the resulting magnetic fields, when scanning the position of the point source, as the Full-Width at Half-Maximum (FWHM) for the depth (in the $z$ direction; computed as the distance away from the tip of the MG) and width (in the $x$ direction; considering the distance from one side to the other side and parallel to the tip of the MG).

\begin{figure}[t!]
     \centering
     \begin{subfigure}[b]{0.49\columnwidth}
         \centering
         \includegraphics[width=\columnwidth]{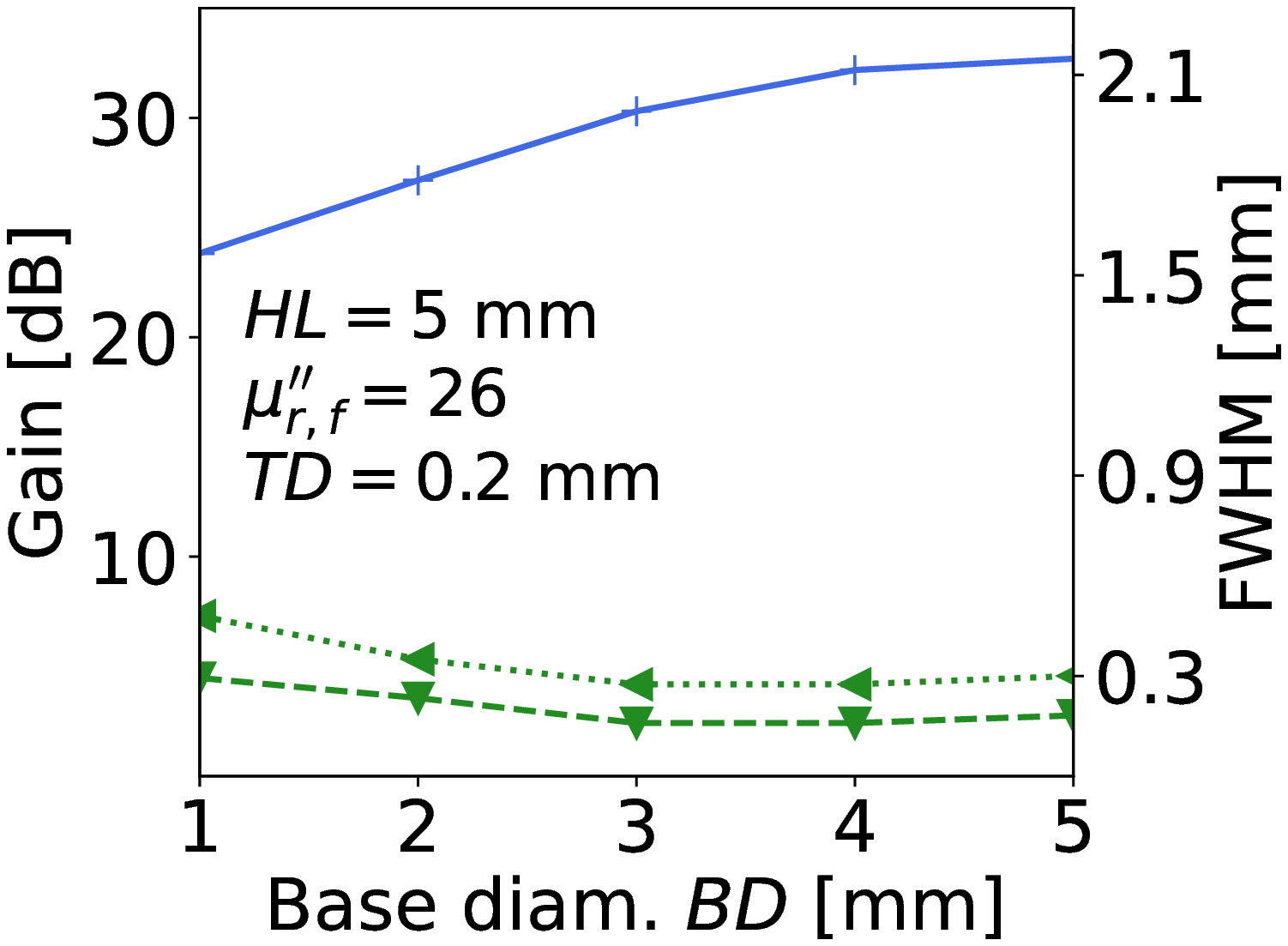}
         \caption{}
         \label{fig:FigTheoreticalSNRcomparisonBRPlotConeSuperanalysis}
     \end{subfigure}
     \hfill
     \begin{subfigure}[b]{0.49\columnwidth}
         \centering
         \includegraphics[width=\columnwidth]{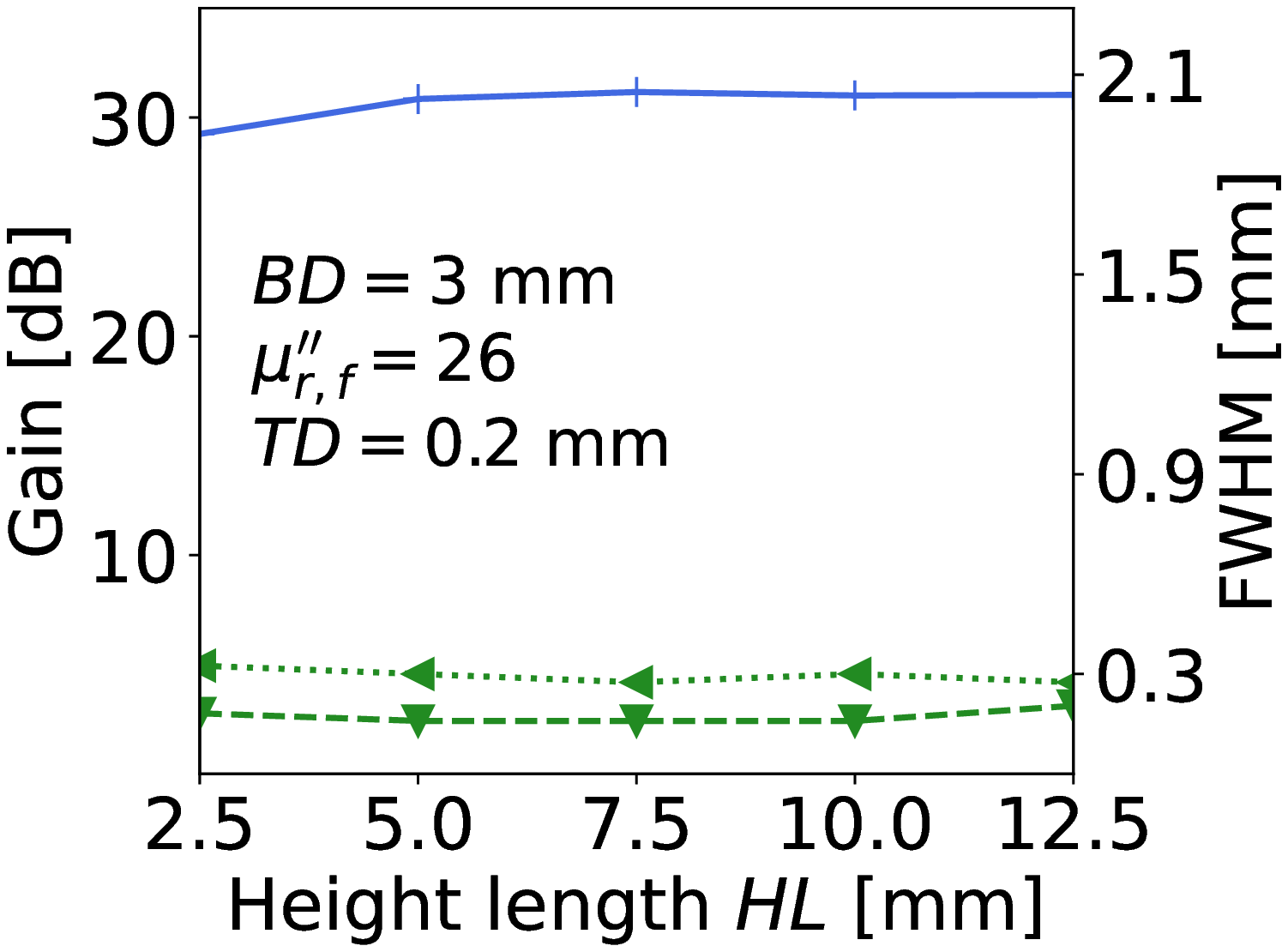}
         \caption{}
         \label{fig:FigTheoreticalSNRcomparisonHLPlotConeSuperanalysis}
     \end{subfigure}
     \hfill
     \begin{subfigure}[b]{0.49\columnwidth}
         \centering
         \includegraphics[width=\columnwidth]{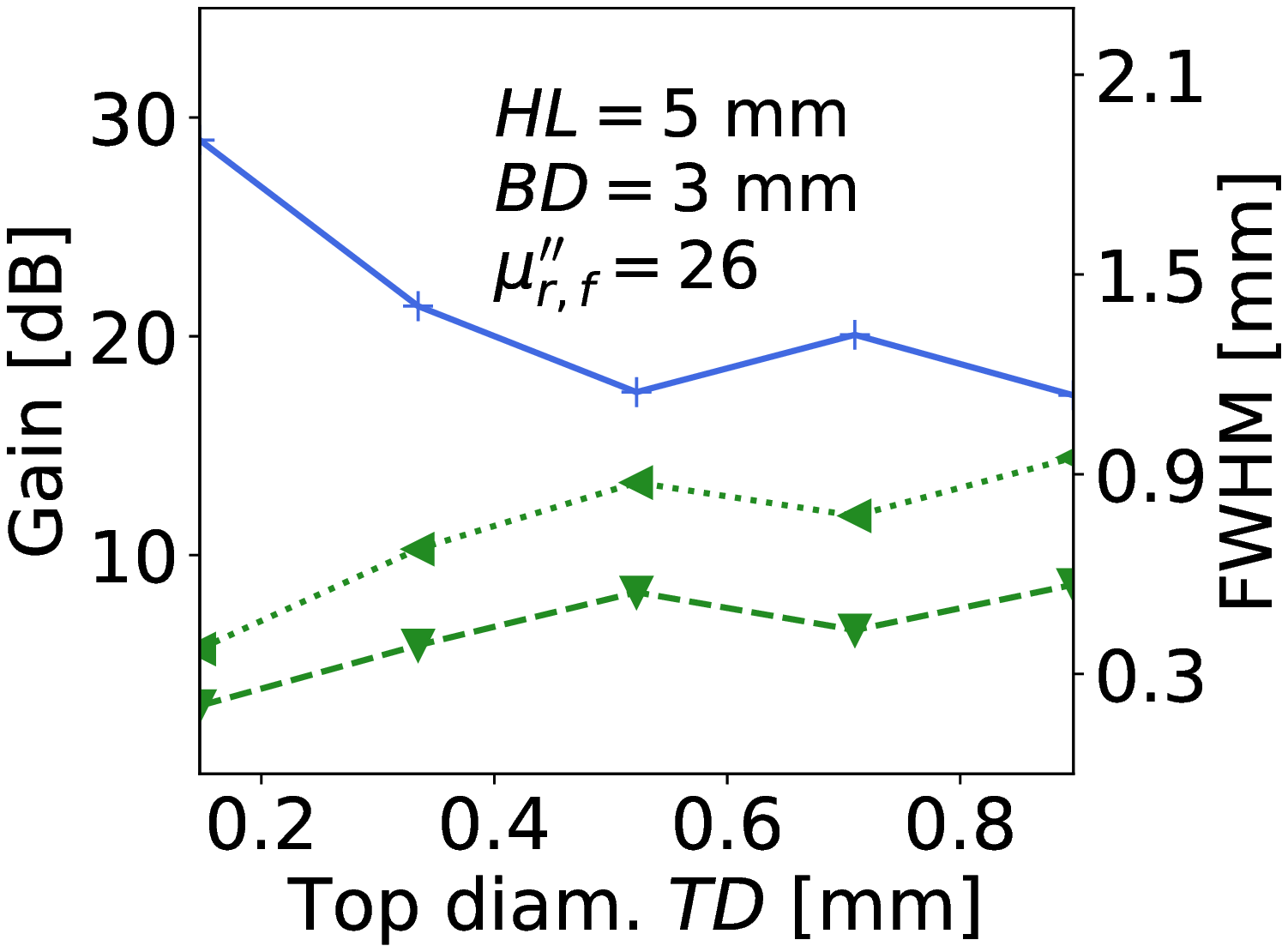}
         \caption{}
         \label{fig:FigTheoreticalSNRcomparisonrTopPlotConeSuperanalysis}
     \end{subfigure}
     \hfill
     \begin{subfigure}[b]{0.49\columnwidth}
         \centering
         \includegraphics[width=\columnwidth]{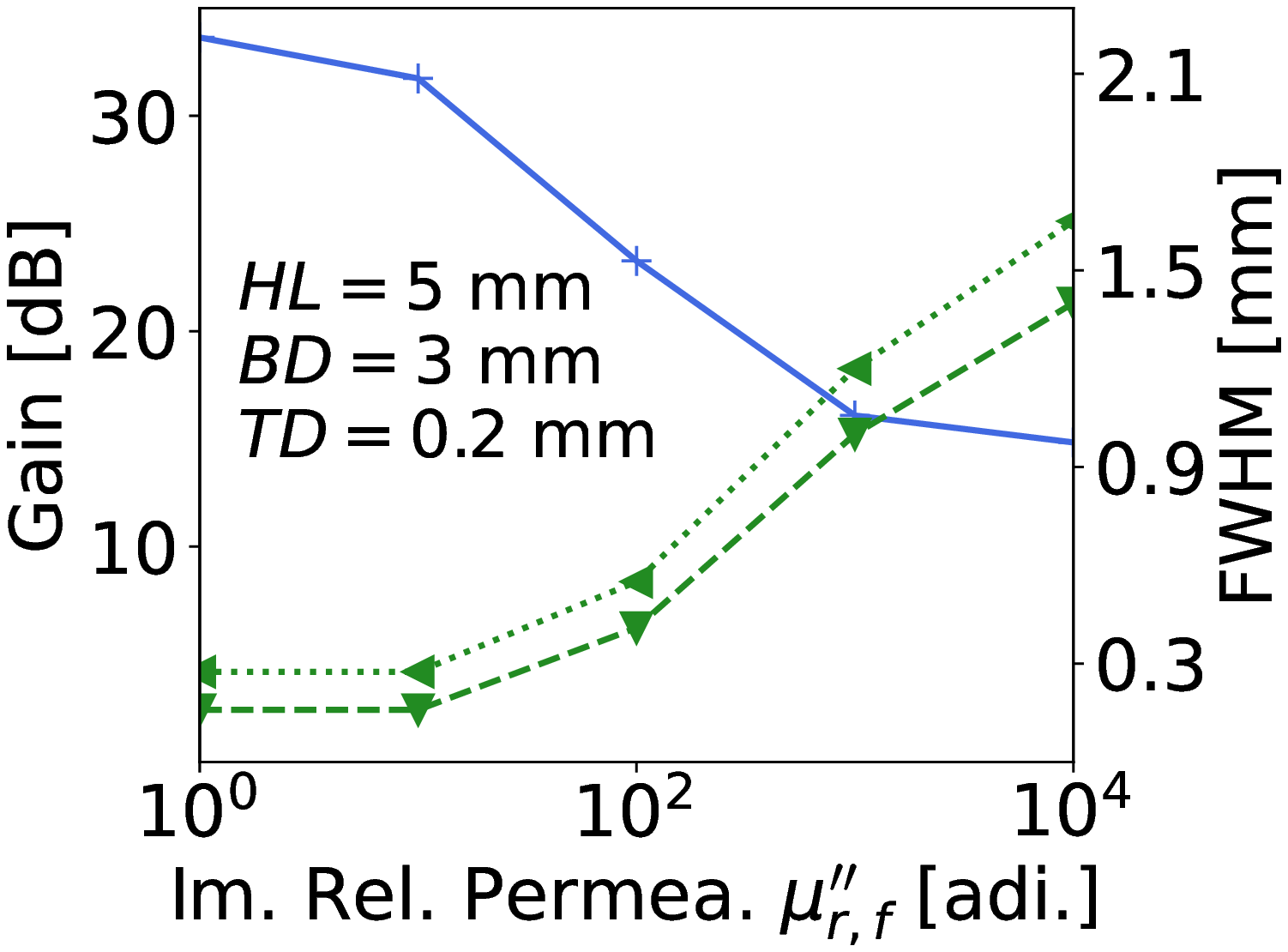}
         \caption{}
         \label{fig:FigTheoreticalSNRcomparisonImRelPlotConeSuperanalysis}
     \end{subfigure}
     \hfill
     \begin{subfigure}[b]{0.1\columnwidth}
         \centering
         \includegraphics[width=\columnwidth]{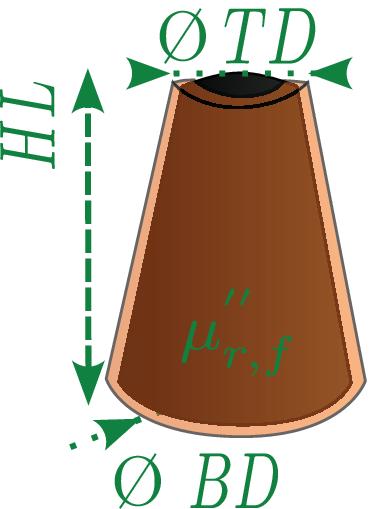}
         \label{fig:SchemeMG}
     \end{subfigure}
     \hfill
     \begin{subfigure}[b]{0.79\columnwidth}
         \centering
         \includegraphics[width=\columnwidth]{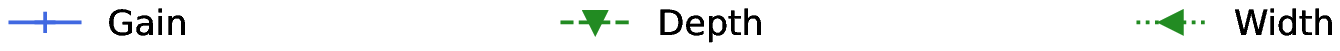}
         \label{fig:FigTheoreticalSNRcomparisonLegend}
     \end{subfigure}
     \vspace{-1\baselineskip}
     \caption{Numerical estimation of the behavior of the MG with regard to magnetic field transferring performance (as gain) and spatial resolution of depth/width (as FWHM) for different parameters.}
    \label{fig:TheoreticalSNRcomparison}
\end{figure}

Then, in one hand, it is shown in Fig. \ref{fig:FigTheoreticalSNRcomparisonBRPlotConeSuperanalysis} that the gain with the MG increases from \SI{24}{\decibel} to \SI{32}{\decibel} with the base diameter ($BD$) of the MG, while its spatial resolution slightly decreases to \SI{300}{\micro\meter}. On the other hand, as shown in Fig. \ref{fig:FigTheoreticalSNRcomparisonHLPlotConeSuperanalysis}, scanning the height length ($HL$) of the MG, the gain increases slightly and the resolution depth and width remain fairly constant around \SI{300}{\micro\meter}. As well, although not shown, as a reference for this analysis, for \SI{5}{\milli\meter} of $HL$, the FWHM (depth and width) without the MG are \SI{1.4}{\milli\meter} and \SI{1.5}{\milli\meter}, respectively.

Furthermore, as proposed in this work, the MG is cone-shaped in order to concentrate the magnetic field of the point source into the OPM, resulting that the smaller the top diameter $TD$, the more gain as shown in Fig. \ref{fig:FigTheoreticalSNRcomparisonrTopPlotConeSuperanalysis}. As well, the spatial resolution curves increase with $TD$, from around \SI{300}{\micro\meter} of spatial resolution to \SI{600}{\micro\meter} and \SI{900}{\micro\meter} for the depth and width, respectively. Therefore, in general, the gain is inversely proportional to the FWHM, both for the depth and width, as in common focusing apparatus (e.g., a focusing lens).

Finally, ferrite materials have losses, and in particular losses at frequencies \SI{<1}{\mega\hertz}~\cite{fairweather_ferrites_1952}. Also, the imperfections of the core surrounded by the superconductor tape encompass leakages, which can be accounted as higher values of the imaginary relative permeability $\mu_{r,f}^{''}$. Therefore, the larger the $\mu_{r,f}^{''}$, the lower the effectiveness of the MG at concentrating and guiding lines of magnetic flux, as shown in Fig. \ref{fig:FigTheoreticalSNRcomparisonImRelPlotConeSuperanalysis}. As a result, with the MG, the gain and spatial resolution degrade strongly for large $\mu_{r,f}^{''}$.
In addition, the losses of the MG produce an effective low frequency noise, which might affect the sensitivity of the OPM, as discussed in detail in Section \ref{SecNoiseEstimation}. Nevertheless, the currently employed materials for the construction of the MG have sufficient proper characteristics in order to demonstrate the capacity of the MG to detect magnetic droplets.

Accordingly, with the analysis of the trends, of the gain and spatial resolution, for the different parameters of the MG, and also due to the mechanical capabilities to assemble the MG, the experimentally used parameters are $BD=3$ mm, $HL=5$ mm and $TD=0.2$ mm (described in Section \ref{DescriptionMG}), whilst $\mu_{r,f}^{''}$ is considered to be $26$~\cite{kim_ultra-sensitive_2016}.

\begin{figure}[t!]
     \centering
     \begin{subfigure}[b]{0.49\columnwidth}
         \centering
         \includegraphics[width=\columnwidth]{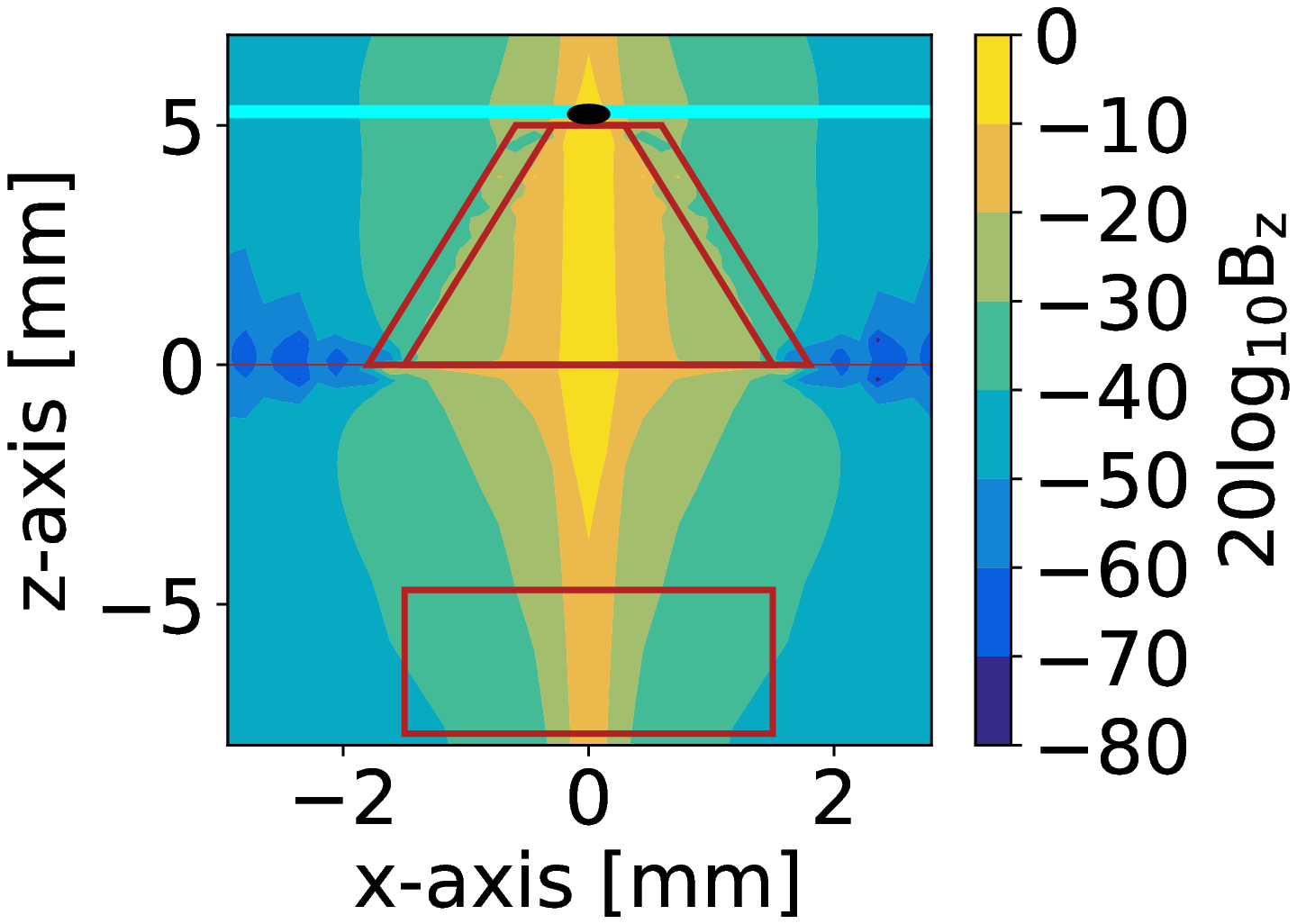}
         \caption{}
         \label{fig:TheoreticalBFieldconeSMnobaseScalarYplanePlot}
     \end{subfigure}
     \hfill
    \begin{subfigure}[b]{0.49\columnwidth}
         \centering
         \includegraphics[width=\columnwidth]{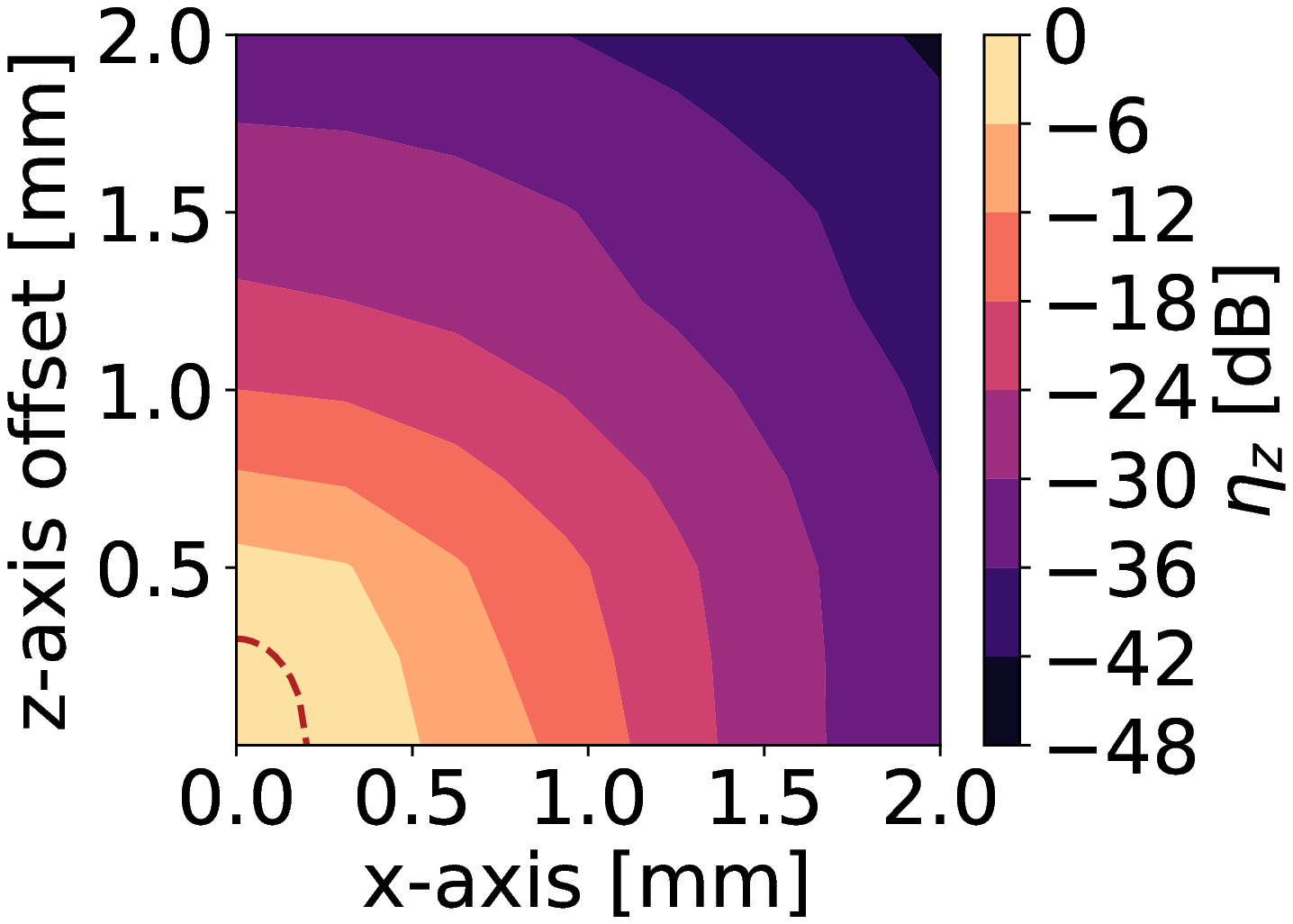}
         \caption{}
         \label{fig:FigTheoreticalFocusInductionconeSuperAnalysis}
     \end{subfigure}
     \caption{Normalized $B_z$ with the OPM-MG system where (a) represents the scalar magnitude and (b) the spatial coupled magnetic field ratio (dashed line indicating the FWHM), both in the $xz$-plane.}
    \label{fig:SchematicsDeviceTheoreticalBfield}
\end{figure}

Subsequently, the theoretical sensed magnetic field of magnetic droplets for the assembled OPM-MG is plotted for the $xz$-plane in Fig. \ref{fig:TheoreticalBFieldconeSMnobaseScalarYplanePlot}, and for illustration purposes it is also represented the microchannel (cyan solid line) extending along the $x$-axis and a magnetic droplet (black circle), as in the experimental setup. Also, from this computed magnetic field, the magnetic field gradient inhomogeneity ratio in the OPM vapor cell, defined in Section \ref{SecGradientSensLoss}, computes to be on average much lower than $1\%$. Hence, this gradient inhomogeneity is negligible compared to the gradient from the compensation coils of the OPM module, as detailed in Section \ref{OPMmodule}.

Finally, for the assembled OPM-MG, in Fig. \ref{fig:FigTheoreticalFocusInductionconeSuperAnalysis} it is plotted the normalized spatial coupled magnetic field ratio $\eta_z$ in the $xz$-plane for differently scanned positions of a magnetic droplet, considered as a point source, with respect to the top-center part of the MG. As expected, it is common to have the maximum coupling ratio at the tip of the MG (top-center part), while the coupling decays as the distance from this point increases. Accordingly, the computed depth/width spatial resolution is around \SI{300}{\micro\meter}, which is appropriate with respect to the dimensions of the microfluidic channel. The latter numerically computed spatial resolution values have been experimentally confirmed as discussed in the next paragraph.

The experimental characterization of the spatial resolution of the OPM-MG system was done during the alignment of the OPM-MG to the microchannel. Accordingly, it was monitored the detected signal level of the magnetic droplets, at a rate \SI{5} drops./s, passing by the tip of the MG while scanning the vertical ($y$-axis) position with the M6 nylon screws (shown in Fig. \ref{fig:MountSetup2}). The maximum signal level was obtained, and then an up-direction and down-direction scan with 1/8 of a turn confirmed that the spatial resolution in the $y$-axis (assuming also to be valid in the $x$-axis, due to the symmetry of the MG in the $xy$-plane) is in the order of one 1/8 of a turn (\SI{125}{\micro\meter} displacement) in either direction, corresponding to around \SI{300}{\micro\meter} spatial resolution in the $xy$-plane. Lastly, the \SI{300}{\micro\meter} numerically computed spatial resolution in the $z$-axis was confirmed by detecting a sufficient signal level from the magnetic droplets when the tip was in total contact with the top surface of the microfluidic chip. At this position, as described in subsection \ref{SecMagDropMicrofluidics}, the distance from the tip to the center of the microchannel is \SI{225}{\micro\meter}, confirming that the $z$-axis spatial resolution is also in the order of \SI{300}{\micro\meter}.

Consequently, a high spatial resolution is achieved with the proposed MG, instead of having a poor spatial resolution similar to the size of the OPM vapor cell, which allows detecting single magnetic droplets, as detailed in Section \ref{SecExperimentalResults}.

\section{Physical magnetometry sensitivity discussion}\label{SecTheoreticalDescription}
For SERF OPMs, optimized for sensitivity and when a magnetic field $B$ changes very slowly, the steady-state solution of the Bloch equation can be used~\cite{seltzer_unshielded_2004,karaulanov_spin-exchange_2016},
\begin{equation}\label{eqn:BlochEquation}
\frac{d \boldsymbol{S}}{dt}=\gamma \boldsymbol{S} \times \boldsymbol{B} + R\left(S_0 \hat x-\boldsymbol{S}\right)-\frac{\boldsymbol{S}}{T_2},
\end{equation}
where $\boldsymbol{S}$ is the spin vector, $t$ is time, $\gamma$ is the gyromagnetic ratio, $\boldsymbol{B}$ is the magnetic field vector, $R$ is the optical pumping rate, $S_0$ is the equilibrium electron spin polarization, $\hat x$ is the direction of optical pumping and $T_2$ is the transverse spin relaxation time.

In this regime, it is more sensitive to use the low frequency modulation signal method, compared to the DC bias method, to optimize the signal response~\cite{savukov_high-sensitivity_2017}. In this case, the response approximately follows $0.77 S_0 \delta B$, where $\delta B$ is the small measured magnetic field; however this is only satisfied with magnetic fields below \SI{200}{\hertz} (while the operation frequency modulation is around \SI{980}{\hertz})~\cite{savukov_high-sensitivity_2017}. Moreover, sensitivity typically is attenuated at magnetic fields higher than $\sim$\SI{50}{\nano\tesla}, which is a higher level than the one for the magnetic droplets considered in this work. Furthermore, at these high levels also magnetic gradients and magnetic materials' noise decrease the sensitivity of the OPM as studied in the following sections.


\subsection{Gradient induced loss}\label{SecGradientSensLoss}
A magnetic field gradient within magnetometers increases the magnetic linewidth of the OPM, leading to a reduction in sensitivity. In particular, when gradients, i.e. inhomogeneities, are present the transversal relaxation decreases as $1/T_{2}=1/T_{2}^{'}+1/T_{AB}$, where $T_{2}^{'}$ is the original transverse relaxation without magnetic field inhomogeneity and $T_{AB}$ is the relaxation caused by the inhomogeneity of the magnetic field~\cite{fang_analysis_2022}. In this work, we can take $T_{2}=T_{2}^{'}(1-\chi)$, where $\chi$ corresponds to the inhomogeneity ratio as $\chi=T_{2}^{'}/(T_{AB}+T_{2}^{'})$. Accordingly, $T_{AB}$ is reduced as the magnetic field inhomogeneity increases~\cite{karaulanov_spin-exchange_2016,savukov_high-sensitivity_2017,song_field_2018}.

Then, assuming a $T^{'}_{2}$ of \SI{5}{\milli\second}, a vapor cell temperature of \SI{150}{\celsius} and a low pressure with a density of atoms of $1.5$ amg (amagat), the transversal relaxation time due to magnetic field inhomogeneity is~\cite{cates_relaxation_1988,fang_analysis_2022}
\begin{equation}\label{eqn:TransRelaxTime}
\frac{1}{T_{AB}}\approx \frac{4 \gamma^2 r_{vc}^4}{175 D}(|\vec{\nabla} B_x |^2 + |\vec{\nabla} B_y |^2 + |\vec{\nabla} B_z |^2),
\end{equation}
where $D$ is the diffusion coefficient, $r_{vc}$ is the equivalent radius of the vapor cell and $|\vec{\nabla} B_x |^2 + |\vec{\nabla} B_y |^2 + |\vec{\nabla} B_z |^2$ is the sum of the squared magnitudes of the gradients for each of the three magnetic field components. Hence, considering the field gradients generated by the magnetic droplets of interest, as discussed in Section \ref{SecSpatialResolutionDiscussion}, $T_{AB}$ remains much larger than $T_{2}^{'}$ and $\chi$ is very small, and the sensitivity is not significantly degraded.

\begin{figure}[t!]
     \centering
     \begin{subfigure}[b]{0.49\columnwidth}
         \centering
         \includegraphics[width=\columnwidth]{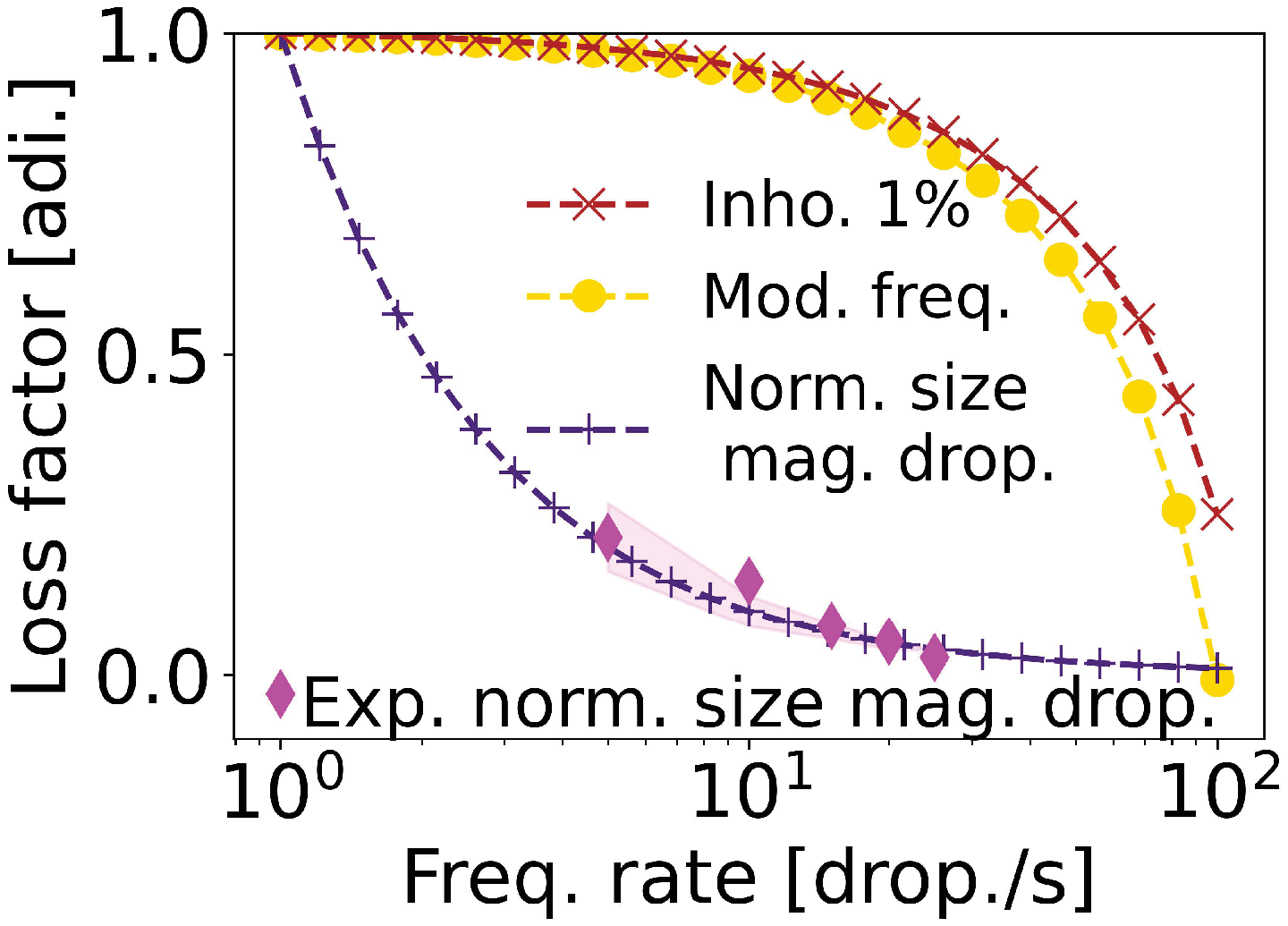}
         \caption{}
         \label{fig:TheorAnalysisSensitivityFreq}
     \end{subfigure}
     \hfill
     \begin{subfigure}[b]{0.49\columnwidth}
         \centering
         \includegraphics[width=\columnwidth]{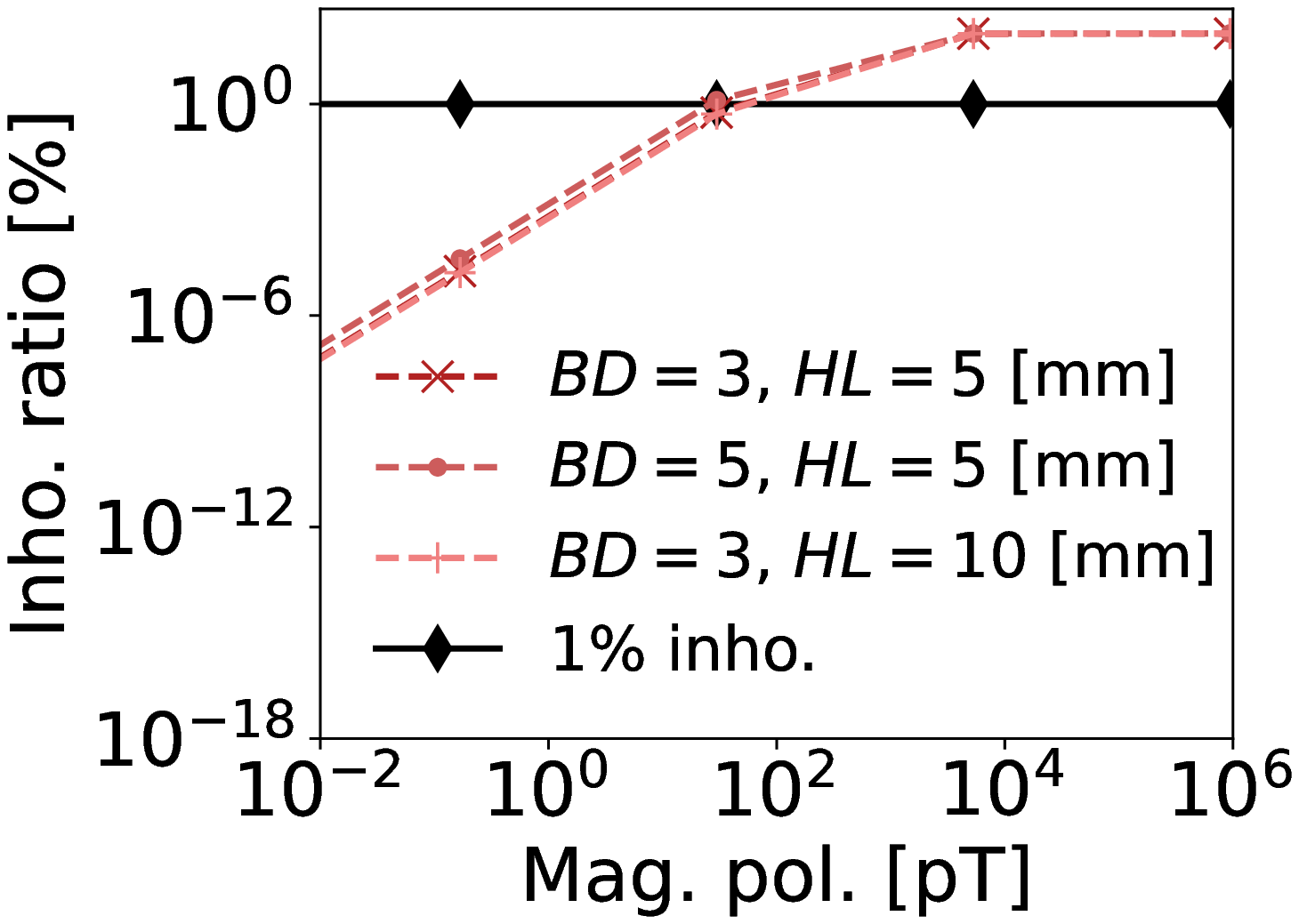}
         \caption{}
         \label{fig:TheorAnalysisSensitivityFreqDifferentInho}
     \end{subfigure}
    \caption{Estimation of the loss factor in sensitivity (or signal) of the OPM module for different rates of droplets, considering (a) $1\%$ magnetic field inhomogeneity ratio, normalized (with respect to \SI{5} drop./s) volume size of magnetic droplets (Norm. size mag. drop.), confirmed with experimental values (magenta diamonds), and \SI{980}{\hertz} modulation frequency (Mod. freq.), and (b) numerical study with different magnetic polarization values, for different dimensions of the MG, to produce a relevant inhomogeneity ratio above $1\%$ (black solid line).}
    \label{fig:SensitivityAnalysisOPM}
\end{figure}

In particular, computing the response of the OPM considering the slightly decreased $T_2$~\cite{savukov_high-sensitivity_2017}, the resultant sensitivity loss factor, for the different rates of the magnetic droplets, is negligible compared to both for the field inhomogeneity generated and for the \SI{980}{\hertz} modulation frequency operating the OPM, as shown in Fig. \ref{fig:TheorAnalysisSensitivityFreq}. Eventually, the effect of magnetic field inhomogeneity does become significant in terms of the loss of sensitivity for strong gradients (high inhomogeneity), as plotted in Fig. \ref{fig:TheorAnalysisSensitivityFreqDifferentInho}. Instead, the size of the magnetic droplets, which is dependent on the droplet rate, is the main factor to the loss of signal within this work, as it reduces with the cubic power of the size of the droplets, as shown in Fig. \ref{fig:TheorAnalysisSensitivityFreq}. In the Figure it is represented the reduction of signal due to the decrease of the volume size of the magnetic droplets, normalized to the maximum volume size of the magnetic droplets at \SI{5} drops./s, over the range of frequency rates studied. Furthermore, this latter trend is confirmed with the experimental detected signal values of the magnetic droplets from Sec. \ref{SecExperimentalResults} (included in the Figure as magenta diamonds).

Hence, relatively small $\chi$ ($<1\%$) is produced by the sensed weak magnetic fields of the magnetic droplets and is not relevant to affect the sensitivity of the OPM. Furthermore, we note that the inhomogeneity ratio scales rapidly with the magnetic polarization to be sensed (determined by the droplet magnetic moment). Instead, the inhomogeneity does not vary significantly for the range of dimensions of parameters of the MG considered, as shown in Fig. \ref{fig:TheorAnalysisSensitivityFreqDifferentInho}. In particular, the magnetic polarization of the magnetic droplets considered is below \SI{50}{\pico\tesla}, thus its inhomogeneity ratio is below $1\%$. Therefore, we notice that the physics of this system, aimed at detecting weak magnetic fields, can support a range of dimensions of the MG to accommodate designs suited for other specific magnetic targets in microfluidic channels.

\subsection{Noise estimation loss}\label{SecNoiseEstimation}
The high permeability magnetic material (ferrite) of the MG generates magnetic field noise, which eventually can limit the sensitivity of the OPM for large dimensions of the MG. Considering the dimensions of the MG described in Section \ref{DescriptionMG}, the level of noise added is computed (supported with numerical computation) to be approximately \SI{265}{\femto\tesla /\sqrt\hertz}~\cite{lee_calculation_2008}, and \SI{300}{\femto\tesla /\sqrt\hertz} experimentally measured as shown in Fig. \ref{fig:FigExpDetectionZFreqDomain} in Section \ref{SecExperimentalResults}.

Even though the compensation coils integrated in the OPM partially compensate for the extra low frequency magnetic field noise in terms of adjusting the operation to have an optimal signal response, the effect on the sensitivity is not avoided. In this direction, the magnetic noise generated by the material of the MG can be accounted as a low frequency baseline level increase, as shown in the Section \ref{SecExperimentalResults}. Also, the noise further increases proportionally to the effective time-of-flight of magnetic droplets. Nevertheless, due to the little time-of-flight of the magnetic droplets and the reduced bandwidth of the OPM used in this work, the loss in sensitivity due to the generated magnetic noise is negligible for detecting the magnetic droplets. 

\section{Characterization of the detected pulses}\label{SecExperimentalResults}
\begin{figure}[t!]
     \centering
     \begin{subfigure}[b]{0.49\columnwidth}
         \centering
         \includegraphics[width=\columnwidth]{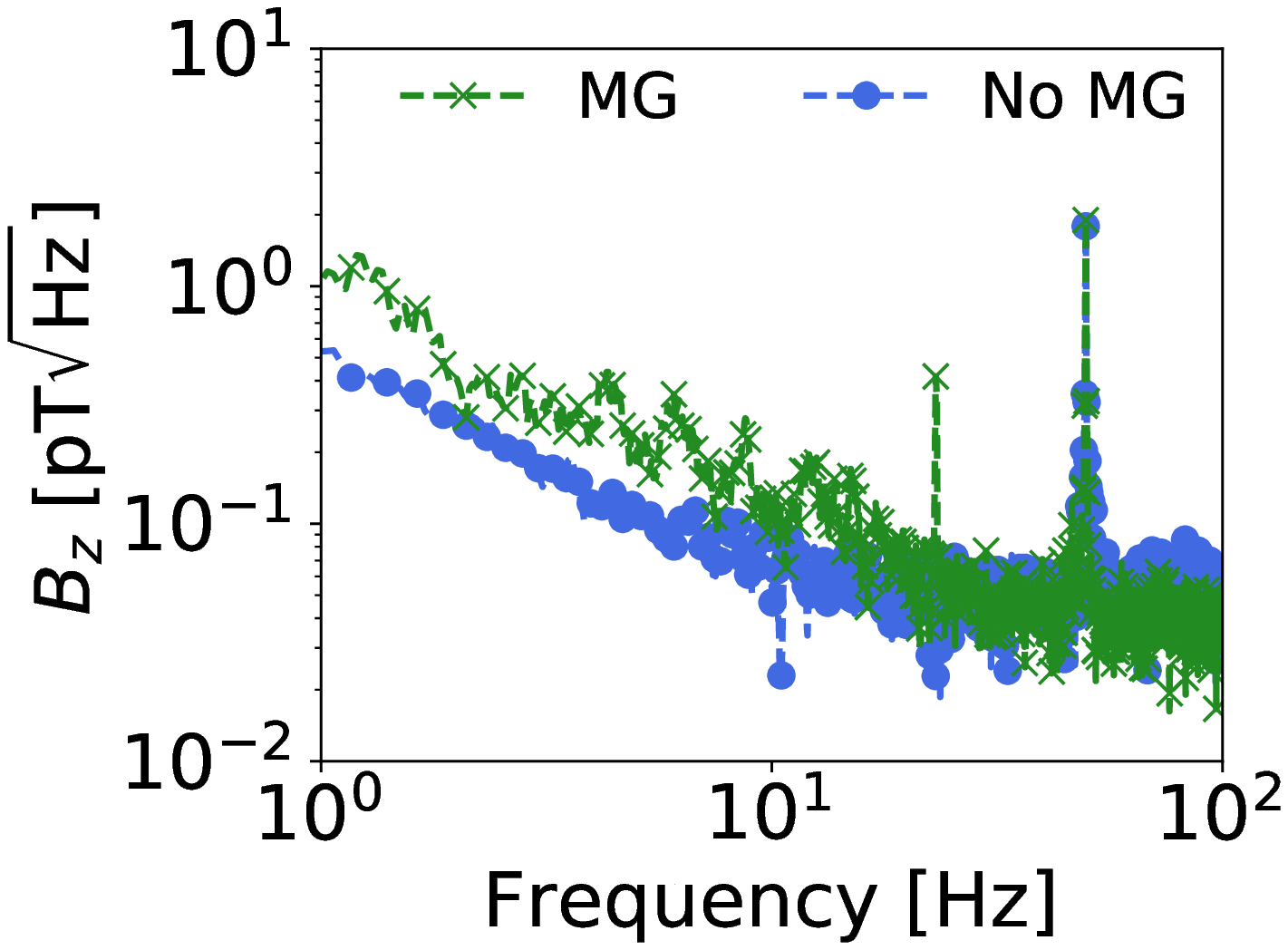}
         \caption{}
         \label{fig:FigExpDetectionZFreqDomain}
     \end{subfigure}
     \hfill
     \begin{subfigure}[b]{0.49\columnwidth}
         \centering
         \includegraphics[width=\columnwidth]{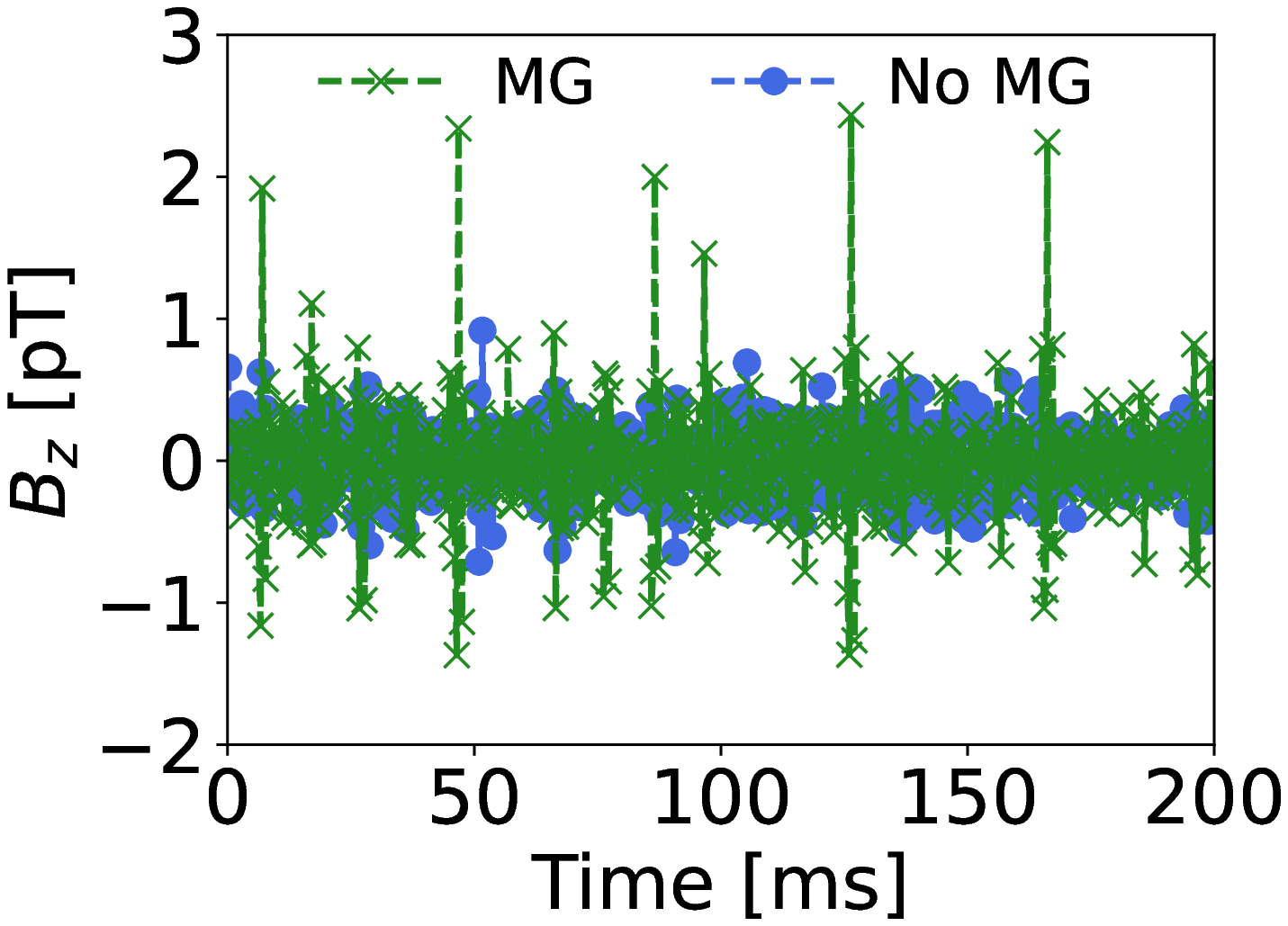}
         \caption{}
         \label{fig:FigExpDetectionFocusZ}
     \end{subfigure}
    \caption{Magnetic droplets detection at a rate of $25$ drop./s (a) in the frequency domain and (b) in the time domain.}
    \label{fig:MagneticDropletsDetection}
\end{figure}

The presented system was adjusted in terms of (i) OPM-MG alignment to the microchannel, (ii) residual magnetic field compensation for the OPM and (iii) microfluidics to generate magnetic droplets.
Firstly, the proper operation of the system was verified by measuring the frequency domain signal of magnetic droplets at $25$ drop./s, as shown in Fig. \ref{fig:FigExpDetectionZFreqDomain}. Initially, the Fast Fourier Transform (FFT), referred to in this article as spectra, was measured without the MG in place. Consistently, no peaks in the signal were present from the magnetic droplets due to the low magnetic flux crossing the OPM. Furthermore, the spectra consistently decays as $1/f$ until the \SI{10}{\hertz} frequency-corner where the baseline becomes flat. A noise peak appears in the spectra at \SI{50}{\hertz} (European power line frequency) due to electronic noise generated by the coils of the OPM, which can be filtered by software.

Then, the spectra with the MG in place was measured, showing a similar trend in the baseline level as the one without MG, but accordingly an extra peak appeared around \SI{25}{\hertz}, matching with the rate of the droplets. Notice that the frequency peak for the droplets stands above the baseline. Furthermore, some extra noise is present in the lower frequencies (below \SI{10}{\hertz}) due to the low frequency equivalent magnetic noise added by the MG, as discussed in Section \ref{SecNoiseEstimation}.

Secondly, in order to characterize different parameters of the detected pulses, time domain measurements were performed for the different rates of droplets, as shown in Fig. \ref{fig:FigExpDetectionFocusZ} for a rate of $25$ drop./s, as an example. Notice that the signal of the OPM was filtered with a second order bandpass filter to \SI{3}-\SI{30}{\hertz} by software, allowing to remove the \SI{50}{\hertz} noise. Due to this signal processing, the peaks are represented without the DC component, thus they show a large positive raise followed by a lower undershoot. In any case, the initial time domain measurement also allowed us to verify that there were peaks detected when the MG was in place, and there were no peaks detected without the MG. 

Then, for the different rates of droplets measured with the OPM-MG system in the spectral and time domain, it was computed the characteristics of the peaks in terms of level, time-of-flight width, noise floor and Signal-to-Noise ratio (SNR), as detailed in Table \ref{tab:FullRange_measurement}. Accordingly, the detected levels of peaks for the different rates of droplets continuously decay with increasing frequency, as expected and shown in the Norm. size mag. drop. curve in Fig. \ref{fig:TheorAnalysisSensitivityFreq}. The peak level of $5$ drop./s is \SI{22.83}{\pico\tesla} and \SI{2.88}{\pico\tesla} at $25$ drop./s. Next, the width of the peaks was computed as the time-of-flight, where the minimum resolvable time-of-flight is around \SI{40}{\milli\second} due to the \SI{300}{\micro\meter} spatial resolution and the \SI{4.5}{\micro\liter} flow rate of the system. As expected, the computed time-of-flights highly follow the width of the generated magnetic droplets for the different rates, from \SI{279.99}{\milli\second} for the lowest rate, down to \SI{36.10}{\milli\second} for the highest rate of droplets considered.

\begin{table}[!ht]
\centering
\caption{$5$ to $25$ drop./s rate characteristics of the detected peaks from the magnetic droplets.}
\label{tab:FullRange_measurement}
\begin{center}
\resizebox{0.89\columnwidth}{!}{
\begin{tabular}{cllll}
\br
\multicolumn{1}{c}{\begin{tabular}[c]{@{}c@{}}Freq. rate\\ {[}drop./s{]}\end{tabular}} & \multicolumn{1}{c}{\begin{tabular}[c]{@{}c@{}}Peak level\\ {[}pT{]}\end{tabular}} & \multicolumn{1}{c}{\begin{tabular}[c]{@{}c@{}}Peak width\\ {[}ms{]}\end{tabular}} & \multicolumn{1}{c}{\begin{tabular}[c]{@{}c@{}}Noise floor\\ {[}$\mathrm{pT/\sqrt{Hz}}${]}\end{tabular}} &
\multicolumn{1}{c}{\begin{tabular}[c]{@{}c@{}}SNR\\ {[}dB{]}\end{tabular}}\\
\mr
$5$ & 22.83$\pm$ 5.64 & 279.99$\pm$ 28.33 & 0.35$\pm$ 0.05 & 21.98$\pm$ 2.18\\ 
$10$ & 15.61$\pm$ 2.59 & 169.13$\pm$ 24.93 & 0.26$\pm$ 0.08 & 21.25$\pm$ 2.60\\ 
$15$ & 8.17$\pm$ 1.31 & 56.92$\pm$ 12.40 & 0.20$\pm$ 0.06 & 17.91$\pm$ 2.54\\ 
$20$ & 5.59$\pm$ 0.68 & 48.43$\pm$ 8.12 & 0.16$\pm$ 0.04 & 16.55$\pm$ 2.13\\ 
$25$ & 2.88$\pm$ 0.08 & 36.10$\pm$ 4.54 & 0.09$\pm$ 0.03 & 15.79$\pm$ 2.50\\ 
\br
\end{tabular}
}
\end{center}
\end{table}

\begin{figure}[t!]
    \centering
    \includegraphics[width=0.49\columnwidth]{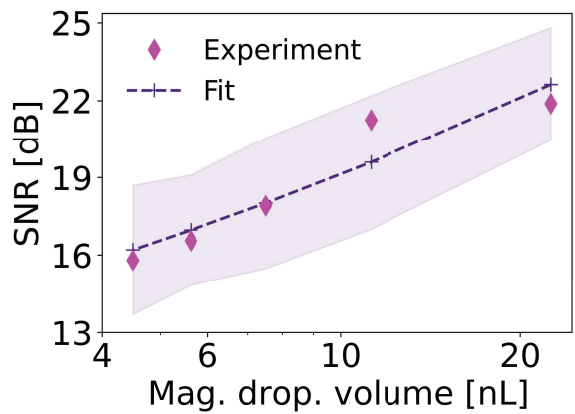}
    \caption{Logarithmic representation of the magnetic droplet volume (in nL) vs. SNR (magenta diamonds) from Table \ref{tab:FullRange_measurement}. Accordingly, an exponential fitting curve (purple dashed line) is also represented to show the near-linear relation with $R^2=0.88$ correlation coefficient.}
    \label{fig:ExpDropSizeSNR}
\end{figure}

In this direction, a sufficient level of detection is a necessary condition for sensing magnetic droplets, but it has to be combined with a measurable variation of the signal when a droplet is sensed. Hence, a more convenient parameter for characterizing the detection of droplets as they flow through the microchannel into the tip of the MG is the SNR ratio; viz. the ratio between the signal, defined as the response magnitude of the peaks detected; over noise, considered as the baseline magnitude variation.

Therefore, the noise floor for the different rates of droplets was quantified in the spectral domain measurements. In particular, for each droplets' frequency rate, the noise floor level was determined as the value at the corresponding frequency, but removing with a Gaussian filter the specific present peak. The obtained values range from \SI{0.35}{\pico\tesla/\sqrt\hertz} to the minimum noise value of \SI{0.09}{\pico\tesla/\sqrt\hertz}, following the trend with the baseline of Fig. \ref{fig:FigExpDetectionZFreqDomain}. In such manner, the SNR for the different rates of droplets are \SI{21.98}{\decibel}, \SI{21.25}{\decibel}, \SI{17.91}{\decibel}, \SI{16.55}{\decibel} and \SI{15.79}{\decibel}, respectively. In general, the SNR obtained for the OPM-MG is above \SI{15}{\decibel} for the range of rates measured. Furthermore, the obtained SNR continuously decays with the drop./s rate, as discussed in subsection \ref{SecGradientSensLoss}, accounting \SI{6.19}{\decibel} reduction for a $5\times$ rate increase. Correspondingly, the rate of droplets defines the volumetric dimension of the magnetic droplets, as smaller droplets for higher rates. Accordingly, the drop./s rate increase computes to a loss factor of $10\log_{10}5=6.99$ dB, and is in agreement with the above SNR decrease. In this direction, in Figure \ref{fig:ExpDropSizeSNR} it is represented, in logarithmic scale, the volume of the magnetic droplets, in nL, compared to the SNR values experimentally obtained. Furthermore, the data has been fitted with an exponential function to compute the linearity (logarithmically) of the trend, achieving a $R^2=0.88$ correlation coefficient. Hence, the obtained SNR follows the expected trend to be logarithmically proportional to the volumetric dimension of the magnetic droplets.

 In this regard, notice that the similarities between magnetic droplets and magnetic signals from cells (e.g., action potentials) can be partially accommodated in size. In particular, for this work, the smaller magnetic droplets are around \SI{100}{\micro\meter} in size, whilst some cells (e.g., large human, Protozoa and Yeast cells) are in the \SI{50}{\micro\meter} diameter range~\cite{tortora_microbiology_2018}. Therefore, when considering the expected \SI{10}{\decibel} signal decrease due to a $\sim 10$x volume reduction of this large cells, the measurements obtained in this work indicate that the detection levels, i.e. of action potential of some large cells, fall within the experimental proven capabilities of the demonstrated system.

\section{Conclusion}\label{SecConclusion}
The use of OPM-MG systems are expected to provide new opportunities, especially with regard to high sensitivity and spatial resolution, in applied sensing physics and other areas of science.
In particular, firstly in this work is described and experimentally demonstrated the possibility of closing the gap between micrometer scale targets and the millimeter dimensions of common OPM by magnetically connecting them with a conical MG. 

Accordingly, the relevant magnetometry physics was presented in this article, with an emphasis on the spatial resolution and sensitivity of an OPM-MG for detecting magnetic droplets in a microfluidic platform.
In this sense, the different quantities measured and computed were discussed, and the role of the MG was highlighted.
Furthermore, it was demonstrated and discussed the capability of the OPM-MG to detect single magnetic droplets in a microchannel with sufficient spatial resolution and SNR.

Moreover, given the high potential of quantum technologies for sensing, we find it crucial to improve the design of magnetic analyzers based on OPMs.
In the coming years, the use of quantum developments in micrometer scale sensing will steadily increase in the form of novel, efficient detecting schemes and also new target sensing, analyzing, and imaging devices.

\section*{Acknowledgments}
We thank C\'esar A. Palacios, Shuhan Wang and Lluis Jofre for comments, support and access to laboratory instrumentation. This work was financially supported by PDR-2014-2022/56-30157-2021-2A grant of the Generalitat de Catalunya and by CICYT PID2019-107885GB-C31 and Prueba de Concepto PDC2022-133091-I00 grants of the Ministerio de Ciencia, Innovaci\'on y Universidades (Spain).

\section*{References}
\bibliographystyle{iopart-num}
\bibliography{2022_MagneticDcPartDet.bbl}

\end{document}